# Using Malware's Self-Defence Mechanism to Harden Defence and Remediation Tools

Jonathan Pan

*Wee Kim Wee School of Communication and Information, Nanyang Technological University, Singapore*

JonathanPan@ntu.edu.sg

*Abstract*— **Malware are becoming a major problem to every individual and organization in the cyber world. They are advancing in sophistication in many ways. Besides their advanced abilities to penetrate and stay evasive against detection and remediation, they have strong resilience mechanisms that are defying all attempts to eradicate them. Malware are also attacking defence of the systems and making them defunct. When defences are brought down, the organisation or individual will lose control over the IT assets and defend against the Malware perpetuators. In order to gain the capability to defend, it is necessary to keep the defences or remediation tools active and not defunct. Given that Malware have proven to be resilient against deployed defences and remediation tools, the proposed research advocates to utilize the techniques used by Malware to harden the tools in a similar manner. In this paper, it is demonstrated that the proposition of using Malware's resilient designs can be applied to harden the tools through experiments.**

## I. INTRODUCTION

Today Malware is a major problem for all nations, organizations and individuals. According to the Organisation for Economic Co-operation and Development (OECD) [1], billions of dollars have been incurred to manage this malice. The McAfee's Quarter 1 2011 report [2] reported that by the end of 2011, they would have gathered 75 million Malware samples. According to a Norton's cybercrime report for 2011 [3], 431 million adults were cybercrime victims in 2010. $338 billion were lost to cybercrime making it a more costly crime relative to drug crime. Malware remains the top threat in such criminal activities with 54 % of the survey respondents citing unfortunate encounters with Malware attacks. It also reported that the number of new Malware released into the wild is continually growing. Malware developers also shifting their focus to mobile smartphone platforms specifically the Android phones which many of us are dependent upon heavily for our daily lives.

There are many anti-Malware products available in the market for individuals or organizations to use to defend against this malice. Defences like Firewall, Intrusion Prevention System (IPS) and Anti-Virus (AV) are intended to protect the IT assets which entail the computing resources, data and supporting infrastructures. These defences are becoming ineffective in both keeping the Malware at bay and contained. AUSCERT reckons that 80% of the Anti-Virus solutions are ineffective in detecting and removing Malware [4]. In a report by MessageLab, it noted that out of 31 Anti-Virus companies, only 6 recognized the malicious file to contain a Malware [5]. According to Yan et al. [6], Malware developers are advancing their software products to enable them to completely bypass the protection of Firewall and Anti-Virus. According to Filiol [30], Malware detection is a NP-hard problem. Besides the fact that these defences are becoming ineffective against Malware, these defences are also becoming victims of attacks by Malware as part of the latter's self-preservation strategy [21], [27]. The starting point of this cyber problem with Malware is their ability to infiltrate. This is typically done through the weakest link in the defence strategy due to humans are involved. One went as far as to comment that it is people's stupidity that led to security lapses [7]. Malware will inevitably infiltrate past the defences and when they do, it is essential to quickly contain these Malware before they induce greater damages to the organizations. According to the Malware Attribute Enumeration and Characterization (or MAEC), which is a Malware knowledge framework defined by MITRE – an American non-profit organization chartered to work on technologies related to public interest, the handling of Malware when the Malware successfully gets pass the defences is known as *Malware Remediation* [8]. However containment or remediation tools used today are not able to match the sophistication of Malware and are being defeated by this malice [17]. The defences and remediation tools will need to stay effective in order to have the ability to defend the IT assets.

This research advocates that the resilient characteristics or resilient design pattern of Malware can be used to harden the defences and remediation tools so that the malice could be negated. The next section of this paper provides an overview of the proposition. A survey of related researches follows. This is then followed by a description of the evaluation and experiments carried out and an analysis of the experimental results. Finally, a conclusion with considerations of the future research directions ends this paper.

## II. MALWARE EPIDEMIC : THE LOSING BATTLE

When a Malware successfully gets past the defences and infects an organization or individual's computing host(s), there will be many risks induced. Such Malware may use the infected computers to steal classified data, launch attacks on other computing resources or disrupt business operations. Control over IT assets is important to both organizations and individuals. The objective of the Malware is to gain control over the IT assets in order to achieve its malicious objectives. The extent of control an organization or individual has over its IT assets following a Malware's successful penetration is proportional affects the risks that the Malware induces to the organization. The following diagram illustrates this premise with a qualitative analysis.

| Security Risks | Organization having More Control | Malware having More Control |
|---|---|---|
| Confidentiality | Reduces data leakage Protects confidentiality | Increases data leakage Losses confidentiality |
| Integrity | Improves security posture Improves reliability of IT assets | Degrades security posture Degrades reliability of IT assets |
| Availability | Improves resiliency Enhances business continuity | Degrades resiliency Affects business continuity |

Table 1: Organizational Risks From Malware Control

In order for the organization or individual to restore control of IT assets, the availability and integrity of defences and remediation tools to be used against Malware are important. When these tools are made defunct, the control of IT assets, in some measures, is gone too. The focus of this research is to study ways to strengthen the resilience of anti-malware solutions and containment tools in fending themselves against the self-preservation attacks of Malware. This aims to enable such solutions or tools to retain their effectiveness to enable the organization or individual to keep or restore control of its IT assets. Before exploring such resiliency designs, the situation confronted by the defence and remediation tools are first studied.

*A. Anti-Malware Solutions: a Victim of Malware ?*

There are many anti-Malware products available in the market. They are intended to protect the IT assets. However they are now becoming victims of Malware attack. Malware that preserves itself by attacking anti-virus software by rendering them defunct are known as *retrovirus* or *retroworm* [9]. They are also known as *"Armored Malware"* [29]. A detailed analysis report on *Conficker* [13] reported that the variant C of the Malware explicitly terminates running security software. Malware are also exploiting the vulnerabilities of anti-virus solution (AV) in order to defunct the defences [26], [27]. According to Landesman [28], since July 2001, ApBot worm was among the first Malware that targeted a range of antivirus software, Trojan detectors and firewall products. Besides eradicating the AV as a means to preserve the Malware, the latter has induced a false negative impression to defences by putting the AV into a 'brain dead' state. According to Vass about Storm Worm [11], this was done to circumvent the network access control (NAC) with quarantine capability that will inhibit unsecure clients from connecting to the network. Anti-Malware companies are developing solutions to protect their products from Malware attacks. However Malware developers are responding accordingly. A question to be considered when developing countermeasures to harden the tools – What can be learnt from the adversary ?

*B. Remediation Tools: Another victim of Malware*

When the Malware gets pass the deployed defences, Malware Remediation is required. This entails using containment processes and tools to facilitate the containment of the raging or defiant Malware. Even then, Malware are being developed to withstand containment or response measures taken against them [17]. These tools used for remediation are being attacked or made defunct by the Malware. An example is the '`W32/Sality.gen.c`' virus (according to Mcafee) [24] that disables the use of Task Manager and Windows registry editors on infected computers. Such tools are required to facilitate Malware remediation. The notorious *Conficker* or '`W32/Conficker.worm`' worm (according to Mcafee) [25] makes explicit attempts to find Malware analysis tools like *wireshark* (network packet monitoring tool), *tcpview* (network packet tool monitoring tool), *procmon* (Sys internals registry monitoring tool) and *gmer* (rootkit detection tool). Malware are also exploiting vulnerabilities in tools used by Malware analysts or incident responders to launch another offensive attack [12].

An obvious reason for losing the effectiveness of the anti-Malware solution and remediation processes or tools to contain them is that the approaches used by the defenders are openly known to all including Malware developers. The latter can then develop mechanisms to overcome or defunct the defences causing the defences to become victims of Malware self-preservation attacks instead of being protectors. Anti-Malware solution developers and incident responders, tasked to salvage Malware infiltration incident, need a means to level off the face-off with Malware developers. The element of known and unknown gave Malware developers an advantage. Defenders need it too.

III. RESEARCH PROPOSITION

The research proposition in this paper is to learn from the resilient techniques or designs incorporated into Malware to preserve themselves and to adopt these techniques or designs into the defensive and remediation tools. The intent is to preserve the availability and integrity of these tools in order to provide the organization and individual with the ability to keep or restore control of its IT assets.

In a study by Alsagoff [10] on Malware's self-protection mechanism, he noted that Malware developers are developing various techniques to enhance the resiliency of their products against any eradication processes. The following table summarizes the qualitative evaluation of the techniques used by Malware developers and their potential reuse by developers of anti-Malware or remediation tools to protect themselves from Malware self-preservation attacks.

| Techniques used in Malware | Relevance to this Research |
|---|---|
| Terminate adversarial and related software (eg, anti-Malware) | Terminate known malicious processes first before they attack |

| Techniques used in Malware | Relevance to this Research |
|---|---|
| Hide or obfuscate Malware files and configuration (including startup) | Hide or obfuscate tool from detection by Malware |
| Protect malicious processes, files and configuration from manipulation | Protect important processes, files and configuration from Malware's manipulation |
| Exploitation of limits of operating system | Exploit limits of operating system to protect tool |
| Disable support tools (eg, task manager, registry editor, startup configuration tool) | Disable support tools that may be used by Malware |
| Redundancy (eg, multiple startup point) | Include similar redundancy capabilities into tool |
| Recovery capabilities (eg, reinstate removed startup point) | Include similar recovery capabilities into tool |

Table 2: Analysis Of Malware Resilience Design

Another prevalent technique used by Malware developers to undermine anti-Malware solutions and Malware analysis attempts is obfuscation [36]. According to Dagon et al. [33], Botnets have considerable resilience to withstand targeted responses against them. They have been known to revive themselves even when they have decapitated [34]. An example (can we be more specific of this example?) of the resiliency of Botnets occurred when researchers analysed, confirmed and initiated the shutdown of a key ISP that hosted the Malware, while the traffic originating from the Malware dropped immediately, it soon recovered [35]. The resilience and robust architecture design of Malware enables the perpetrator(s) to keep its offensive effectiveness against its targeted victims and to fend off response measures by the victims' defenders.

Malware developers are incorporating advance technologies to enhance the resiliency of their products. Malware like *Zellome* incorporated Artificial Intelligence capabilities to facilitate *polymorphic behaviour* to protect itself from being detected [22]. Malware also have anti-forensic capabilities to prevent any forensic or analysis attempts to be done on them in order to contain them [9], [12]. Brand et. al [12] cited various anti-forensics mechanisms that have been incorporated into modern Malware such as anti-online analysis which prevent online analysis from being able to analyse the Malware, anti-dissemblers to prevent reverse engineering the Malware, and, anti-tools which exploit vulnerabilities to analysis tools applied to them. Other notable attributes of Malware advancement include self-preservation, self-healing and self-updating capabilities. Such attributes have been incorporated to enhance the resilience against attempts to eradicate them. According to Shevchenko [23], there are various forms of self-preserving capabilities being incorporated into Malware. The following diagram illustrates the approaches used. They are namely in the dimensions of generic or targeted against countermeasures used by defence or containment tools, also whether the self-preservation approaches are passive or active in nature.

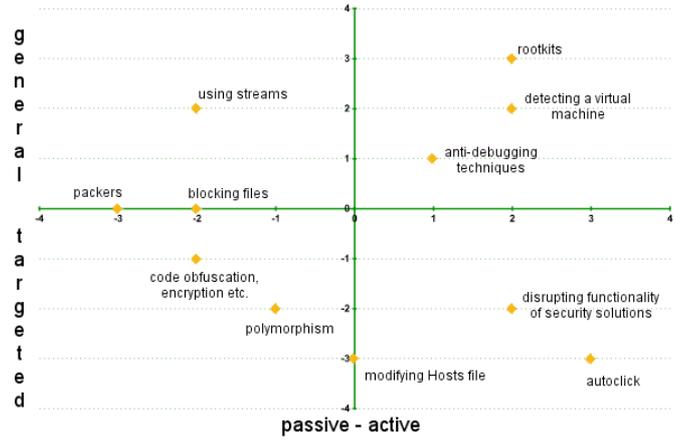

Figure 1: Extract from Shevchenko [23]

Malware has been known to launch counter offense that includes destruction of the infected system when they detect an attempt to take them down is being initiated [32].

IV. RELATED WORK

There are a number of researches done to harden or protect defences or remediation tools after attacks. According to Xue [26], there are vulnerabilities in anti-virus software. The author noted four kinds of vulnerabilities in such software. Firstly, they are local privilege escalation where the antivirus software like other software is at risk of being exploited to gain uncompromised access to the Operating System. Secondly, ActiveX vulnerabilities have been exploited. Antivirus software that uses such technology is exposed. Thirdly the engine of the antivirus software is complex due to the complexity of the adversary that it is trying to defeat. However with such complexity, there will be vulnerabilities in them which can be exploited. Finally, most antivirus software has management components to facilitate the administration of such software. Typically they use client server designs to develop these components. Such design uses various forms of TCP/IP communication protocols to communicate which in turn have vulnerabilities. Xue's recommendation to mitigate the risks of vulnerability exploitation in antivirus software was to adopt security development lifecycle into the development of antivirus software, and to conduct audits and fuzzing tests on the software products. Finally, it was recommended to setup an avenue for updates to be pushed down in a timely manner. Another researcher, Srinivasan, advocated that antivirus software can be protected against Malware attacks by hiding the antivirus software from other processes including those belonging to the Malware [27]. His solution entails changing the names of the files, registry entries and using a different name. His proposition matches that of the behaviour of Malware. However he stopped short of stating the techniques used by Malware. Kerivan and Brothers [31] conducted a series of attack tests, likened to those initiated by Malware, against a few security software namely *IPS* and *antivirus*. Their test demonstrated that none of the security software could fend themselves against such forms of attacks, however the *IPS* was more robust than the antivirus software.

However, none of the above mentioned researches explicitly studied Malware's resiliency designs or mechanisms and assessed whether they can be incorporated into defence tools used in prevention and containment to enhance its resilience. This forms the motivation and proposal of this study.

V. METHODOLOGY

In order to verify that the hypothesis that Malware's self-preservation techniques can be used in tools for defence or remediation, three experiments were done and the results were analysed to assess their effectiveness – that is to remain resiliently active when Malware like offensive attacks are used against these tools. The independent variable is the resilient design technique used in the tools. The dependent variable is the specific attribute to which the resiliency design is seeking to protect or harden against Malware attacks. In the experiments that were carried out, (A) a specific attribute is selected where the resilient characteristics of the Malware will be applied upon or built around and where the focus of an attack vector that is typically used by Malware will be applied against. The specific attribute used in this experiment would be either a dependent software like the firewall of the targeted host, registry setting, or the availability of the tool itself. (B) The attack vector was chosen from Mitre's *Common Attack Pattern Enumeration and Classification (CAPEC)* [14] as it represents a body of knowledge on attack patterns. CAPEC is used by *Malware Attribute Enumeration and Characterization* or *MAEC* [15] which the latter defines the behaviour and characteristics of Malware while the former defines the Malware's offensive behaviours. (C) An application was developed that codified a specific resilient techniques used by Malware to protect a specific attribute. The application was developed using Visual Express C# 2010. Prior to the start of the experiment, (D) a check is first done that to ensure that the attribute to be preserved for the test is set or running well. This will be the pre-test step for the experiment. (E) A tool that mimics the Malware's offensive behaviour to attack software was used to carry out the self-preservation attacks. The offensive tool used in our experiments was Metasploit Framework Version 4.0.1. This tool has been used to develop Malware or offensive hacking [18], [19], [20]. (F) The application, which was developed to protect the attribute, would, either pre-emptively or be triggered, to preserve or protect the attribute when the Metasploit tool is used. The activation of the preservative measures by our custom built software could be triggered manually or automatically. Finally (G) a verification was done to check whether the attribute to protect was effectively preserved against the assault. This will be the post-test step for the experiment. As Metasploit Framework's Meterpreter was used in the experiment, the resident protection of the antivirus installed in the test host was disabled to facilitate the use of the former (former or formal?) tool to carry out offensive attacks against the test application.

The setup of the experiment was done in a virtualization environment running on a Mac Pro laptop. Virtualization was done using Virtualbox version 3.2.1. Two guest operating systems were used. One was the targeted host, which ran on Windows XP SP3, from which the application ran and where the attribute to protect resided. This host was then connected by Meterpreter to the other virtualized host running BackTrack 5 with Metasploit Framework Version 4.0.1 included. Throughout the execution of the experiments, Internet connectivity was not enabled and the experiments were carried in an isolated environment.

A. *Experiment 1*

The objective of the experiment is to show that a dependent software like firewall can be protected using self-preservation techniques used by Malware. The attribute in this experiment is the firewall service. This attack vector used is to turn off the firewall which is also known to CAPEC as CAPEC-56: Removing/short-circuiting 'guard logic'. The following are the pseudo code for the application used.

```
Create command line instruction to start firewall using "netsh"
CALL System Procedure to start new process with command line
   instruction
Return Error code of command line instruction
```

Figure 2: Pseudo Code To Restore Disabled Service

The following are the outputs of the experiment to show that initially the firewall was working with no warning notification raised by the Operating System, this was followed by the attack vector to disable the firewall done through Metasploit's Meterpreter and finally re-enable the firewall by the proposed application. The proposed application carried out the re-instatement of firewall by clicking on the button that corresponded to the test. The following diagram is the application GUI prior to the start of the experiment.

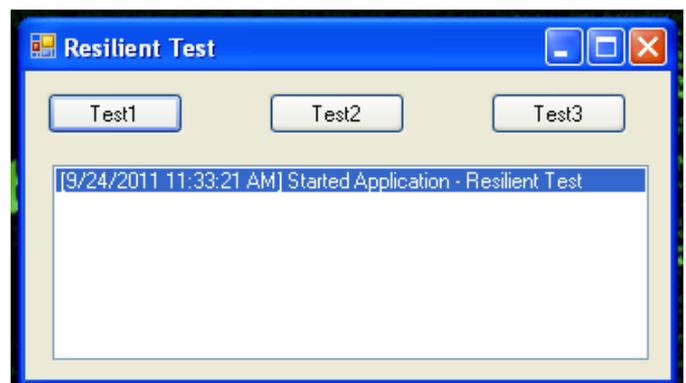

Figure 3: Snap-Shot Of Custom Developed Application For Experiment

The following notification alert by the Operating System indicated that only the AV was not running. The firewall is running still.

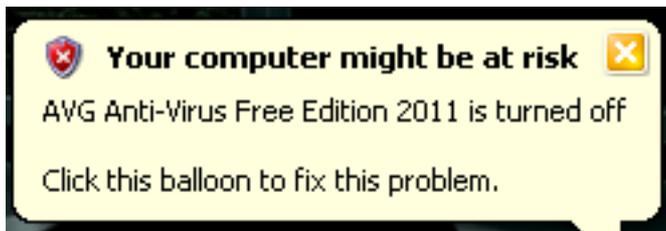

Figure 4: Security Alert Message That Anti-Virus Was Turned Off

Next, the offensive attack was initiated with a firewall termination instruction from Metasploit, the targeted host reported that the firewall is not running as shown below.

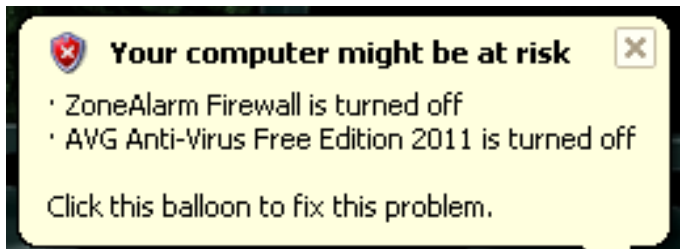

Figure 5: Security Alert Message That Firewall And Anti-Virus Were Turned Off

The application, running on the targeted host, was instructed to reinstate the firewall service on the targeted host.

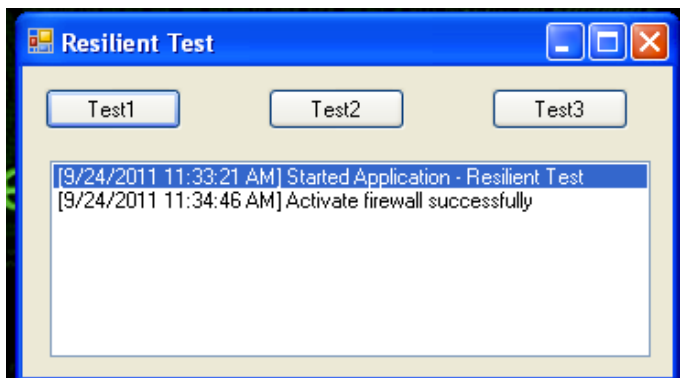

Figure 6: Status Reflected On Custom Developed Application

(F) The following is the response notification alert from the targeted host showed that the firewall was back on again indicating our software successfully restored the status of the firewall.

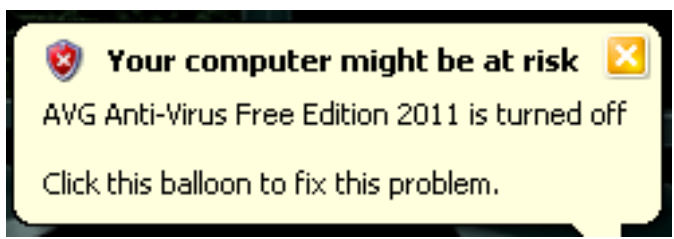

Figure 7: Alert Status Message Indicated That Only Anti-Virus Was Turned Off

### B. *Experiment 2*

The objective is to demonstrate the use of self-preservation techniques to preserve registry settings from undesired changes. The technique used to preserve the registry setting is to constantly monitor the setting and to respond when a change (update or delete) occurs. In this experiment, the attack vector is to induce a change to the registry setting flag value. This attack vector is also known as CAPEC-203: Manipulate Application Registry Values. The followings are the pseudo code for the application that have been developed to preserve the registry setting flag value.

```
// Check registry setting is created; applicable for initial run
IF Registry key NOT EXIST THEN
    Create Registry Key
SET Registry Key with required flag value
END IF

// Start polling sequence to monitor the registry
FOR 5 iterations for this experiment
    GET Registry Key value

    // For scenario when registry key is removed
    IF Registry Key NOT EXIST THEN
      Create Registry Key
      SET Registry Key with required flag value
    END IF

    // For scenario when registry key is changed
    IF Registry Key <> desired value THEN
      SET Registry Key with desired value
    END IF

    PAUSE for 2 seconds
END FOR
```

Figure 8: Pseudo Code For Protecting Registry Settings

The following are the outputs of the experiment to show the stages of the tests, pre-experiment registry setting value, the registry change executed by Metasploit's Meterpreter and finally restoration of registry value by the proposed application.

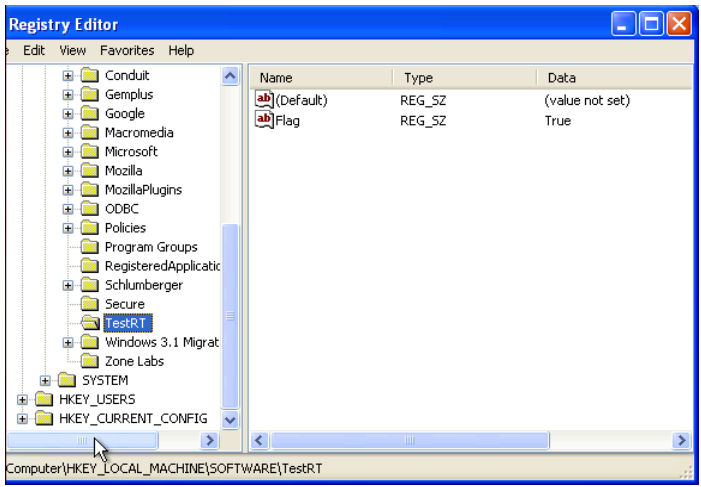

Figure 9: Flag Status (True) Prior To Start Of Experiment

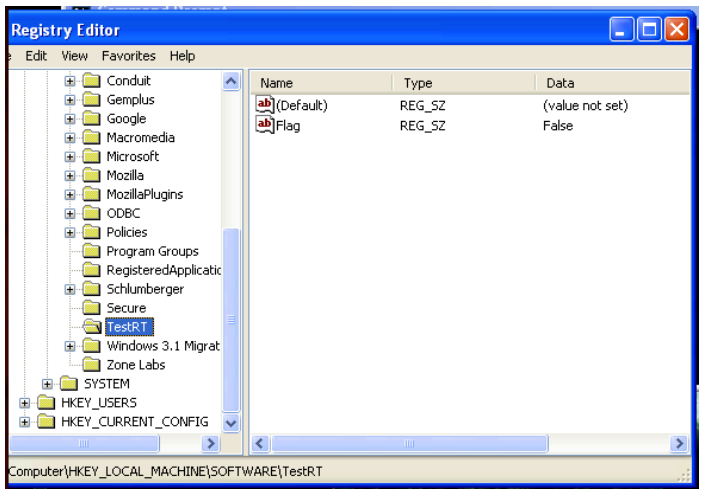

Figure 10: Changed Induced To Registry Settings

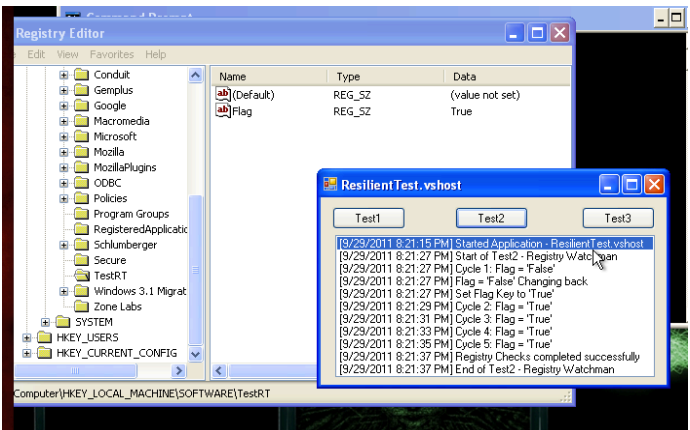

Figure 11: Custom Application Restored Value Of Registry Settings

## C. Experiment 3

The objective of this experiment is to demonstrate the use of self-preservation techniques to preserve the application that we developed ourselves from offensive termination attack. In this experiment, the offensive attack carried out was to terminate our application. This attack vector is also known as CAPEC-17: Accessing, Modifying or Executing Executable Files. The self-preservation technique used is to randomize the process name of the application and executable file so that it cannot be detected by the offensive software, Metasploit, hence protecting the application from sudden abrupt termination. The following is the pseudo code of the application.

```
GET Current application path details which includes directory path and file name
CALL Generate New Randomized Application Name
CALL Copy Original Application File with a New Application File Name in same directory path
Call Application Exit function for original application process
```

Figure 12: Pseudo Code For Obfuscation Of Custom Application Process Name

The followings are outputs from the experiment. The outputs entail Metasploit successfully identified and terminated our application, the application randomized its process name and Metasploit was not able to identify and terminate the application.

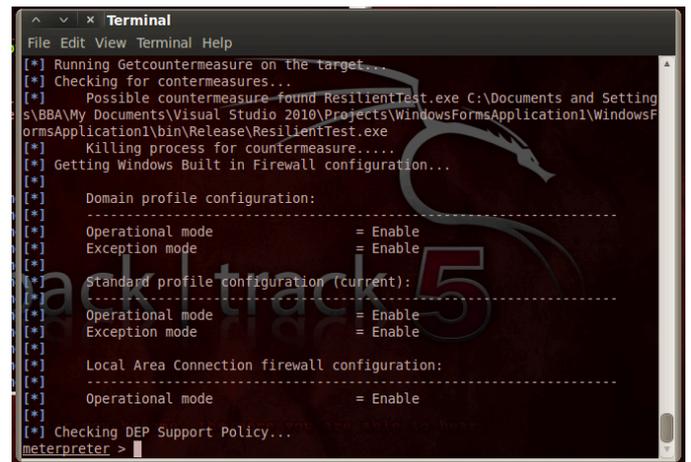

Figure 13: Use Of Metasploit To Identify And Terminate Targeted Application Process

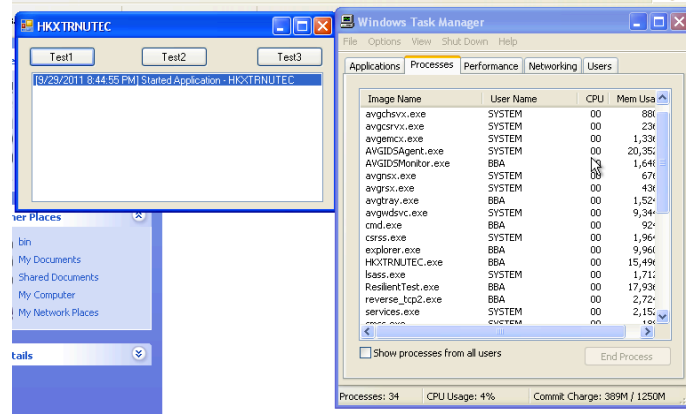

Figure 14: Custom Application Used Randomly Generated Process Name

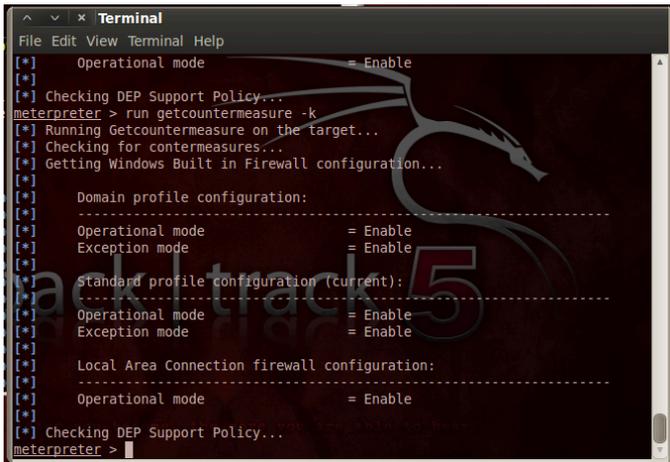

Figure 15: Metasploit Could Not Identify And Terminate The Application

*D. Analysis Of Results*

The experiments demonstrated that the Malware's resilience software design techniques could be codified and incorporated to protect the applications. These and other Malware's resilient design approaches can be applied to security defence or remediation tools in order to harden them against self-preservation attacks of Malware.

VI. DISCUSSION OF FINDINGS

Resilient design techniques used in Malware can be incorporated into the security tools, however there are some considerations as to how these techniques should be applied. The following are considerations gathered from the experiments.

- What attribute to be protected
- Why resilient technique is applicable
- When should such resilient techniques be applied
- Where to apply resilient design technique
- Which resilient design technique is relevant
- How to implement such resilient techniques
- Self-preserving offensive techniques by Malware
- Residual risks when using such technique

For security defense and remediation tools to protect the organization or individual from being infiltrated by the extensive variety of Malware or the ever changing behavior of the Malware, more than one of the resilient techniques may be required to be applied. A knowledge or rule based technique will be required to learn about the attack and deploy the appropriate counter measures accordingly.

VII. CONCLUSION & FUTURE DIRECTIONS

Malware are attacking security solutions in order to preserve themselves for longer periods and in turn to induce more risks and damages to organizations and individuals. The current situation is in favor of the Malware and their developers. The proposition in this paper is to learn from the Malware characteristics in the way they harden their products and to apply their resiliency design techniques to the security solutions. Such resilient design approaches can be formalized into a body of knowledge in the form of design patterns. They can then be subsequently incorporated into defense or containment/remediation tools in order to prolong the effectiveness of the tool much like the intent of Malware. In addition, the attack vectors of the Malware can be used to evaluate the resilience strength of the tool as part of the security development lifecycle.

Future research options to this proposition are:
a. Conduct experiments with real Malware to further verify the research proposition.
b. Mapping of Malware attack vectors to resilient designs.